# Fast Neutron Detector for Fusion Reactor KSTAR Using Stilbene Scintillator

**Seung Kyu Lee[a], Byoung-Hwi Kang[a], Gi-Dong Kim[b], and Yong-Kyun Kim[a*]**

[a] *Department of Nuclear Engineering, Hanyang University,*
*17 Haengdang-dong, Seongdong-Gu, Seoul, 133-791, Korea*
[b] *Ion Beam Application Group, Korea Institute of Geoscience and Mineral Resources,*
*Daejeon, 305-350, Korea*
*E-mail:* ykkim4@hanyang.ac.kr

ABSTRACT: Various neutron diagnostic tools are used in fusion reactors to evaluate different aspects of plasma performance, such as fusion power, power density, ion temperature, fast ion energy, and their spatial distributions. The stilbene scintillator has been proposed for use as a neutron diagnostic system to measure the characteristics of neutrons from the Korea Superconducting Tokamak Advanced Research (KSTAR) fusion reactor. Specially designed electronics are necessary to measure fast neutron spectra with high radiation from a gamma-ray background. The signals from neutrons and gamma-rays are discriminated by the digital charge pulse shape discrimination (PSD) method, which uses total to partial charge ratio analysis. The signals are digitized by a flash analog-to-digital convertor (FADC). To evaluate the performance of the fabricated stilbene neutron diagnostic system, the efficiency of 10 mm soft-iron magnetic shielding and the detection efficiency of fast neutrons were tested experimentally using a $^{252}$Cf neutron source. In the results, the designed and fabricated stilbene neutron diagnostic system performed well in discriminating neutrons from gamma-rays under the high magnetic field conditions during KSTAR operation. Fast neutrons of 2.45 MeV were effectively measured and evaluated during the 2011 KSTAR campaign.

---

[*] Corresponding author.

# Contents

## 1. Introduction

Nowadays, measuring fast neutrons is the most crucial issue in facilities such as particle accelerators, generation-IV nuclear reactors (GEN-IV), and nuclear fusion reactors (KSTAR in Korea). Neutron diagnostics is one of the most important tools for controlling burning plasmas in future fusion devices such as the Korea Superconducting Tokamak Advanced Research (KSTAR). Various neutron diagnostic tools are used in a fusion reactor to measure plasma parameters such as the fusion power, power density, ion temperature, fast ion energy, and their spatial distributions. The stilbene scintillator is proposed for use as a neutron-flux monitor to measure the characteristics of neutron radiation in the KSTAR fusion reactor.

## 2. Composition of neutron diagnostic system

### 2.1 Design of neutron detector system

The fast neutron diagnostic system is designed to detect wanted-directional fast neutrons, resist strong magnetic fields, and reduce high background levels of accompanying gamma-rays. The schematic of the neutron diagnostic system array is illustrated in figure 1. The collimator utilizes polyethylene assemblies to shield against neutrons out of the sight line. Lead reduces gamma-ray generated by neutrons incident to the polyethylene and a backscattered gamma-ray from the external structure of KSTAR device. The size of the collimator tube is a 50 mm diameter for the aperture and 400 mm length [1]. The housing of the neutron detector is surrounded by 10-mm thick soft magnetic iron to remove the influence of the magnetic field [2].

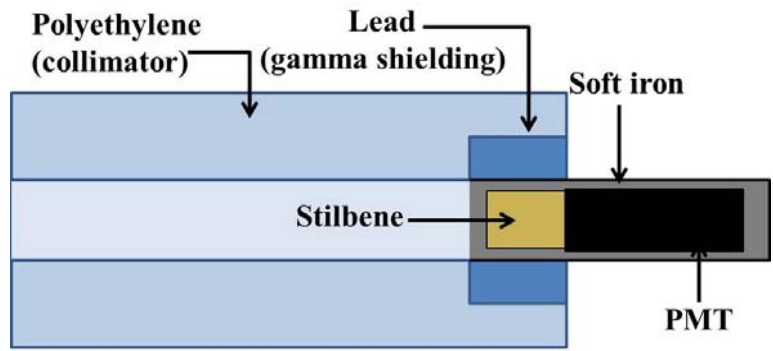

**Figure 1.** Structure of one-channel neutron detector and its collimation geometry.

## 2.2 Stilbene crystal scintillator

A stilbene is an organic scintillator crystal ($C_{14}H_{12}$) that is widely used for fast neutron detection. The neutron flux and spectrum measurement are based on proton recoil detection in the stilbene crystal. Single crystal stilbene scintillators have been used for many years as a representative of fast neutron detectors in the range of 500 keV to 20 MeV due to their good timing performance (<10 ns), detection efficiency, and excellent pulse shape discrimination (PSD) properties in mixed neutrons and gamma radiation fields [3]. It also makes it possible to perform neutron field spectrometry and estimate the neutron-yield evolution with a high temporal resolution against the background of accompanying gamma-rays in experiments with non-stationary neutron sources. The scintillator demonstrates very high capability for neutron-gamma pulse shape discrimination and can provide high gamma-ray suppression that is essential for fast neutron spectrometry. A stilbene scintillator's light output efficiency is low: about 30% of that of a NaI(Tl) scintillator's. In contrast, a stilbene scintillator is more advantageous when the light output created by charged particles and electrons is compared [4], [5]. Due to this advantage, stilbene crystal scintillators are used widely to measure neutrons and discriminate them from the accompanying gamma-rays in the background.

A stilbene crystal scintillator was specially developed for neutron measurements during the KSTAR campaign. It was grown using the Bridgeman-Stockbager method [6], [7]; typical properties are presented in table 1.

**Table 1.** General properties of stilbene crystal [8]

| Property | Value |
|---|---|
| Light output (%, anthracene) | 50 |
| Density (g/cm$^3$) | 1.16 |
| Decay time (ns) | 4.5 |
| Peak Emission (nm) | 410 |

## 2.3 Digital signal processing

When measuring fast neutron spectra on a gamma-ray background, specially designed electronics is necessary. The signals of neutrons and gamma-rays are discriminated in real time by a pulse shape discrimination circuit that is implemented into the flash analog-to-digital convertor (FADC). The analog signals are directly fed into the FADC and digitalized. The analyzed data are transferred to an online monitoring PC via a 2 MB Ethernet.

The neutron and gamma-ray signals are discriminated by the digital charge discrimination method (digital charge comparison method, DCC) [9]. The DCC method uses the total to partial charge ratio analysis, as shown in figure 2 [3]. The total charge (Q) is integrated from the peak of the signal pulse ($t_0$) to the end where the signal pulse returns to the baseline ($t_2$). The partial charge ($\Delta Q$) is integrated from some delay time (from $t_0$ to $t_1$) after the peak of the pulse to ends ($t_2$). The FADC is equipped with a sampling frequency of 400 MHz and vertical resolution of 10 bits.

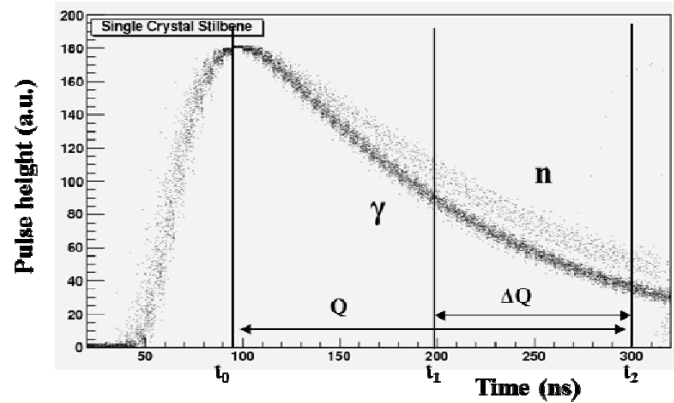

**Figure 2.** n-γ separation with digital charge comparison (DCC) method. Sample pulses were obtained by using a 200 MHz flash ADC.

## 3. Experiments

### 3.1 Calibration of stilbene neutron detector using gamma-ray source

Neutron and gamma-ray sources were used to check the operation of the stilbene neutron detector to verify its energy linearity and neutron-gamma pulse shape discrimination (PSD) properties.

For the first experiments, the energy for the gamma radiation was calibrated. A Φ 50 mm × 40 mm stilbene scintillator was used. This stilbene scintillator was attached to a 2-inch 7195 Hamamatsu photomultiplier tube (PMT) with optical grease (BC-630, Saint-Gobain) on the PMT's window. Figure 3 shows a schematic diagram of the experimental setup for the radiation measurements. A high voltage of 1700 V was applied to the PMT. Figure 4 shows the calibration results for the stilbene scintillator. For the given energy range, the response of the stilbene scintillator was linear for the gamma radiation. All of the measurements were performed under the same conditions. $^{54}$Mn, $^{60}$Co, and $^{137}$Cs gamma sources were used in the energy calibration. Scintillation signals were proportional to the energy of the corresponding center of the Compton edge; this was defined as the middle point from the peak to the Compton edge, as shown in figure 4.

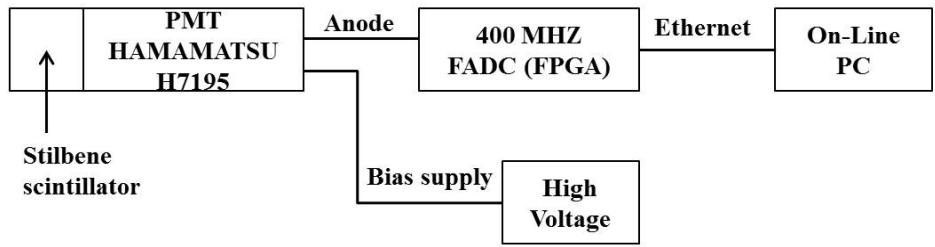

**Figure 3.** Schematic diagram of data acquisition system for measuring fast neutron responses.

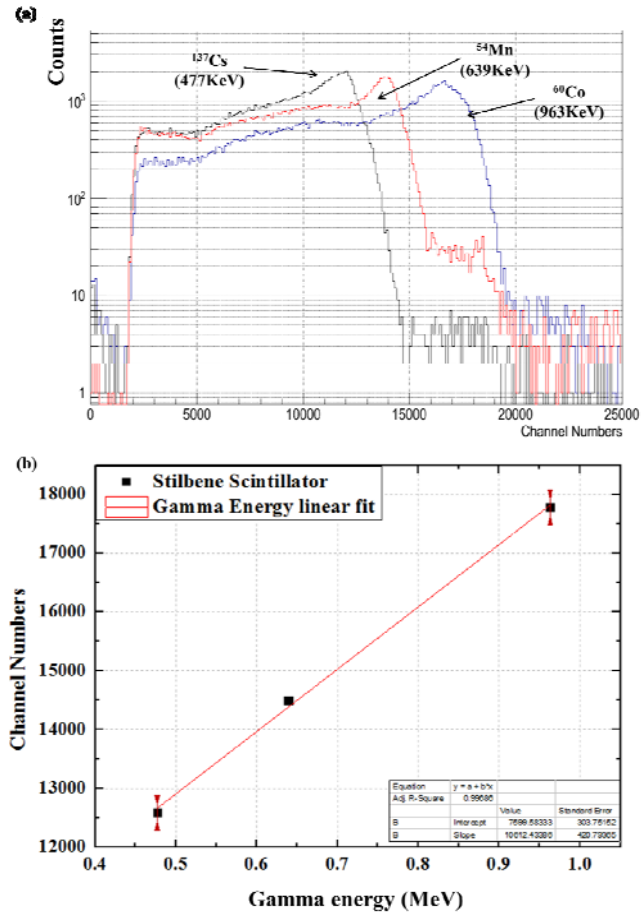

**Figure 4. (a)** Gamma-ray pulse height spectra for photon energy calibration of stilbene scintillator; **(b)** calibration results of amplitude for measurement setup in scale.

The correlation coefficient R is the square root of $R^2$, which is a quantity that shows how good the fit is; it can be computed by the following equation:

$$R^2 = \frac{Explained\ \ variation}{Total\ \ variation} = \frac{TSS\ -\ RSS}{TSS} \tag{3.1}$$

where TSS is the total sum of the square and RSS is the residual sum of the square. SD is the root mean square of the error or the standard deviation of the model [7]. In the results, the R value of this stilbene was 0.997; it showed good linearity for gamma radiation.

– 4 –

### 3.2 Performance of neutron/gamma-ray separation using neutron radiation source

A 33-µCi $^{252}$Cf neutron radiation source was used for n-γ separation experiment of a stilbene scintillator detector. The experimental setup was the same as that in the above experiments, as shown in figure 3. The neutron and gamma-ray signals were discriminated by using a digital charge PSD method that uses total to partial charge comparison, as shown in figure 2.

Figure 5(a) shows the two-dimensional scattered plot of the measured partial charge versus total charge of the stilbene scintillator with a 33-µCi $^{252}$Cf neutron fission source. The upper parts come from neutron interactions, and the lower parts are from gamma interaction since both neutrons and gamma-ray are emitted from the source. For the electron-equivalent energy ($E_{ee}$) calibration of the same detector, two gamma-ray sources of $^{137}$Cs and $^{60}$Co were used. Figure 5 (b) shows the particle identification distribution (PID) at 500 keVee, which was obtained from figure 5(a) and is defined by following equation:

$$PID = \frac{\Delta E - f_\gamma(E)}{f_n(E) + f_\gamma(E)} \tag{3.2}$$

where $f_\gamma$ and $f_n$ are fitting function of gamma-rays and neutrons to $E_{ee}$. To evaluate the separation of the stilbene scintillator, a figure-of-merit (FOM) was used and defined as:

$$FOM = \frac{S_{n\gamma}}{FWHM(n) + FWHM(\gamma)} \tag{3.3}$$

where $S_{n\gamma}$ is the distance between the centroids of the neutron peak and gamma-ray peak in the PID spectrum, as shown in figure 5(b) [3]. The n/γ separation property, or FOM of the stilbene scintillator, was evaluated to be 1.40 at 500 keV$_{ee}$. Finally, we obtained recoil proton spectra generated by fast neutrons of $^{252}$Cf, as shown in figure 5(c). Curve 1 in figure 5(c) shows the total recoil protons and recoil electrons spectrum of a stilbene scintillator irradiated with $^{252}$Cf, and curve 2 shows only the recoil proton spectrum.

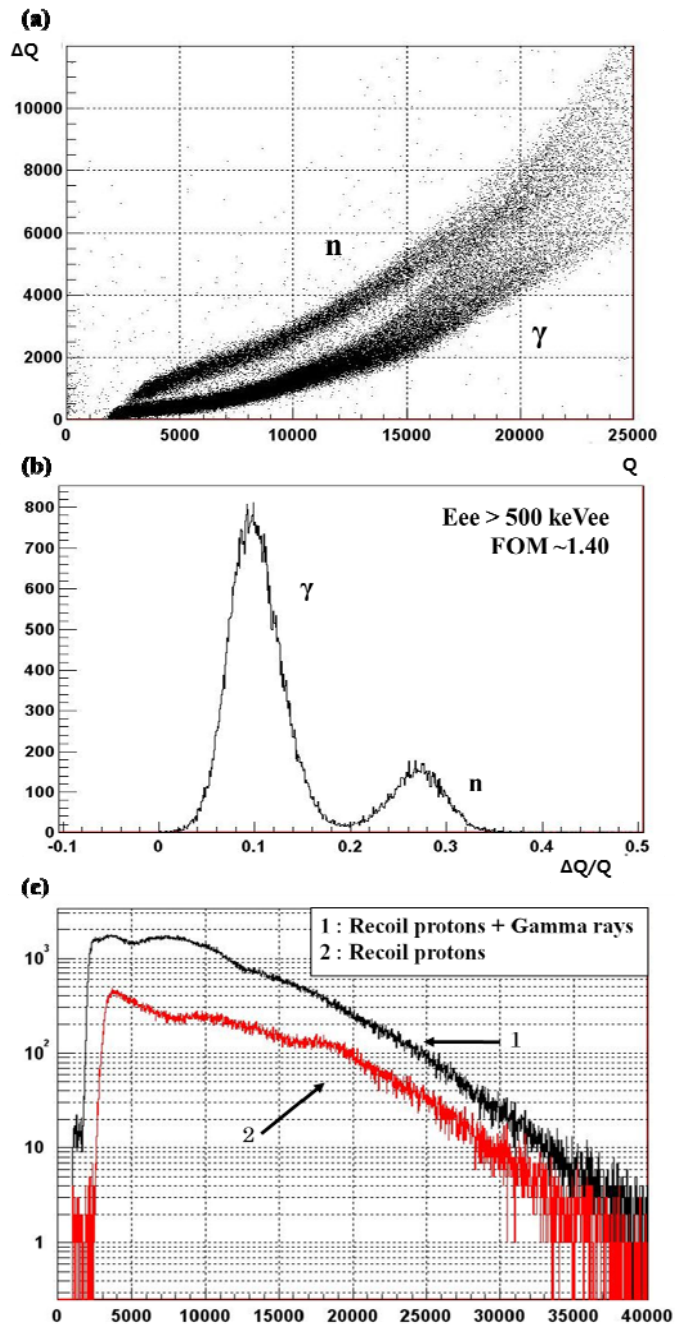

**Figure 5.** (a) Neutron/gamma-ray scattered plot by digital charge comparison method for $^{252}$Cf neutron source of stilbene scintillator, (b) neutron/gamma-ray particle identification distribution (PID) in stilbene scintillator, and (c) recoil proton spectrum obtained with $^{252}$Cf source of stilbene scintillator.

### 3.3 Calibration of stilbene neutron detector using mono-energetic neutrons

Mono-energetic fast neutrons were produced by the Korea Institute of Geoscience and Mineral Resources (KIGAM) using a thin target film and protons through the (p, n) reaction. Neutron energies of 1.5, 2.0, and 2.5 MeV were produced, and the energy spread width was about 300 keV. These experiments were performed under the same conditions as the above experiments, and detectors were located 380 cm from the neutron source.

Figure 6(a) shows the recoil proton spectrum, and the end point corresponding to the maximum energy of the (p, n) reaction was clearly observed for the stilbene scintillator. Figure 6(b) shows the function of recoil proton energies to have some linearity. The correlation coefficient R can be calculated using equation (3.1). The R value of this stilbene was 0.996; there was a good linearity performance for the mono-energetic neutrons.

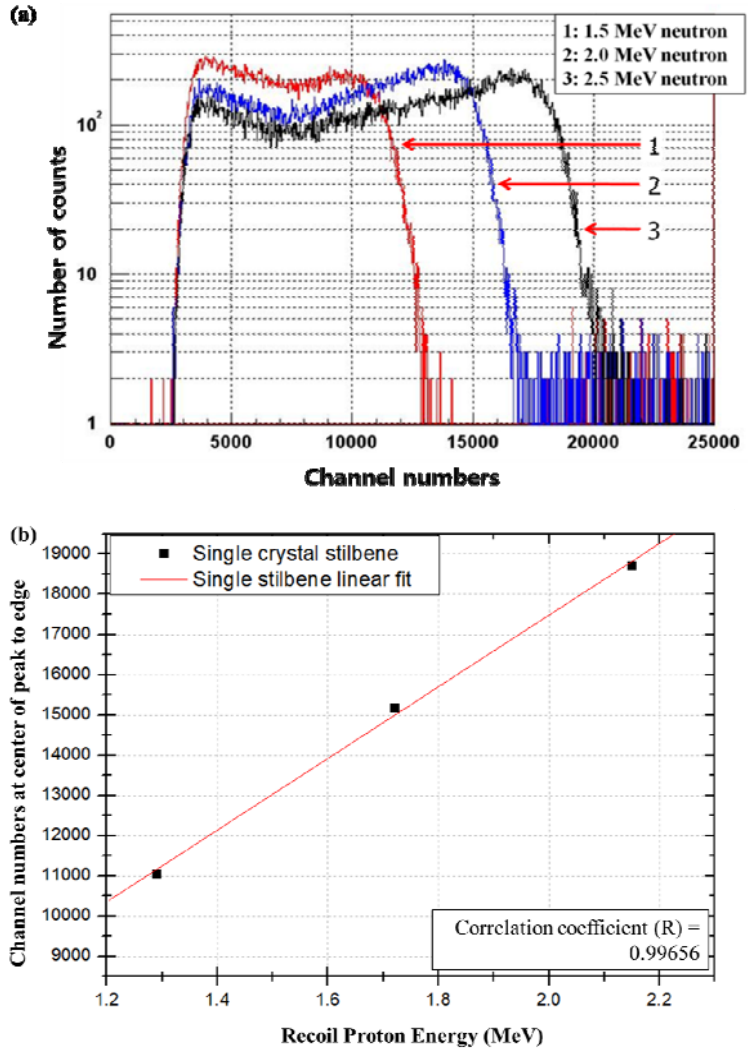

**Figure 6.** (a) Response function of stilbene scintillator for mono-energetic fast neutrons; (b) result after calibrating the amplitude for measurement setup in scale.

### 3.4 Relative intrinsic efficiency for an angular dependence

A 33 μCi $^{252}$Cf neutron radiation source was used for these experiments using a stilbene scintillator detector. The angular dependence experiment was performed to measure the intrinsic efficiency according to the direction of the incident angle in the polyethylene collimator. The polyethylene collimator shields against neutrons out of the sight line. To confirm the shielding performance of the polyethylene, an experiment to count neutrons coming from different angles was needed. The experiments were performed by positioning the neutron source at 0°, 30°, 45°, 60°, 90° relative to the neutron detector and 40 cm from the stilbene scintillator at each angle. Figure 7 shows the stilbene detector system and the experimental setup for the angular dependence.

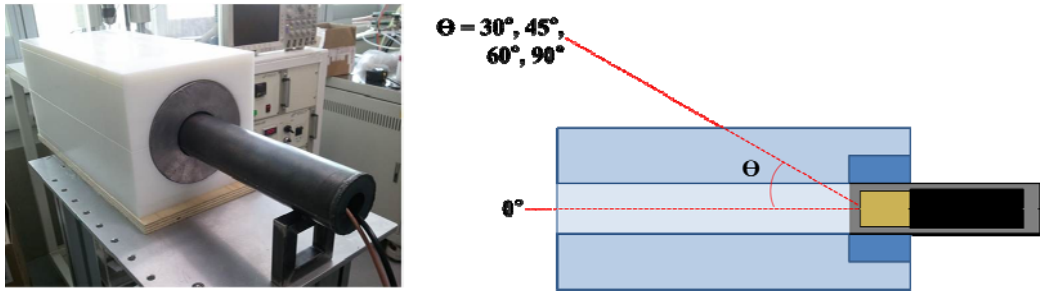

**Figure 7.** Stilbene neutron detector system and experimental setup for angular dependence.

The intrinsic efficiency of fast neutron detection can be calculated by the following equation (3.4):

$$Instrinsic\ efficiency = \frac{Number\ of\ pulse\ reccord}{Number\ of\ neutron\ incident\ on\ detector} \tag{3.4}$$

Table 2 shows the total number of fast neutron counts for the stilbene scintillator at each degree measured. The calculated value of the shielding efficiency was over 50%. Figure 8 shows the relative detection efficiency spectrum with respect to 0°.

| Degree | 0° | 90° | 60° | 45° | 30° |
|---|---|---|---|---|---|
| Counts (Neutrons) | 20637 | 10277 | 6213 | 4082 | 1680 |
| Intrinsic efficiency | 0.0193 | 0.0096 | 0.0058 | 0.0038 | 0.0016 |
| Relative efficiency | 1 | 0.4977 | 0.3012 | 0.1980 | 0.0814 |

**Table 2.** Neutron detection efficiency for angular dependence.

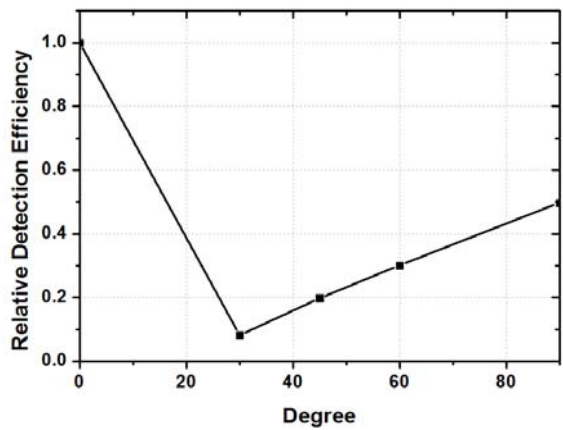

**Figure 8.** Relative detection efficiency spectrum for each degree with respect to 0°.

As the angle decreases, the neutron path length of the collimator grows longer. Thus, the number of detected counts and the intrinsic efficiency were confirmed to be lower. Thus, the shielding efficiency of this diagnostic system was estimated to be 50%.

### 3.5 Relative intrinsic efficiency for magnetic shielding

These experiments were performed under the same conditions as the above experiments. The PMT associated with the neutron diagnostic system will be located in the fringe field of the solenoid at KSTAR. The central field of the solenoid will be about 3.5 T. A magnetic field simulation indicated a maximum fringe field of approximately 150 gauss of PMT. We examined the effect of magnetic shielding of the phototube photocathode using neodymium magnets (1000 G, 4000 G). A detector assembly consisting of a H7195 PMT and soft iron magnetic shield with 10 mm thickness was used. The experimental setup for the magnetic shielding is shown in figure 9.

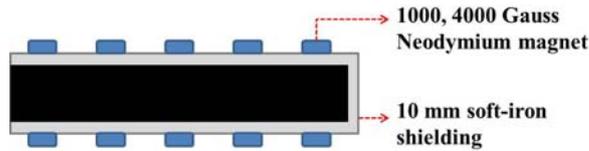

**Figure 9.** Experimental setup for magnetic shielding using 10-mm-thick iron.

The effect of the magnetic shielding is shown in table 3 and figure 6. The relative detection efficiency of the neutron detector using a magnetic shielding of 10 mm thick soft iron was above 1000 Gauss magnetic field. The relative efficiency of the neutron diagnostic system using a magnetic shielding of 10 mm thick soft iron was above 97% and 80% relative detection efficiency in the 1000 and 4000 Gauss magnetic fields, respectively.

|  | No magnet, without magnetic shielding | 1000 G, without magnetic shielding | 1000 G, with 10 mm magnetic shielding | 4000 G, with 10 mm magnetic shielding |
|---|---|---|---|---|
| Counts (neutrons) | 13467 | 11 | 13204 | 10833 |
| Intrinsic efficiency | 0.012 | ~ 0 | 1.85 | 0.015 |
| Relative efficiency | 1 | ~ 0 | 0.98 | 0.80 |

**Table 3.** Neutron intrinsic detection efficiency for soft iron magnetic shielding.

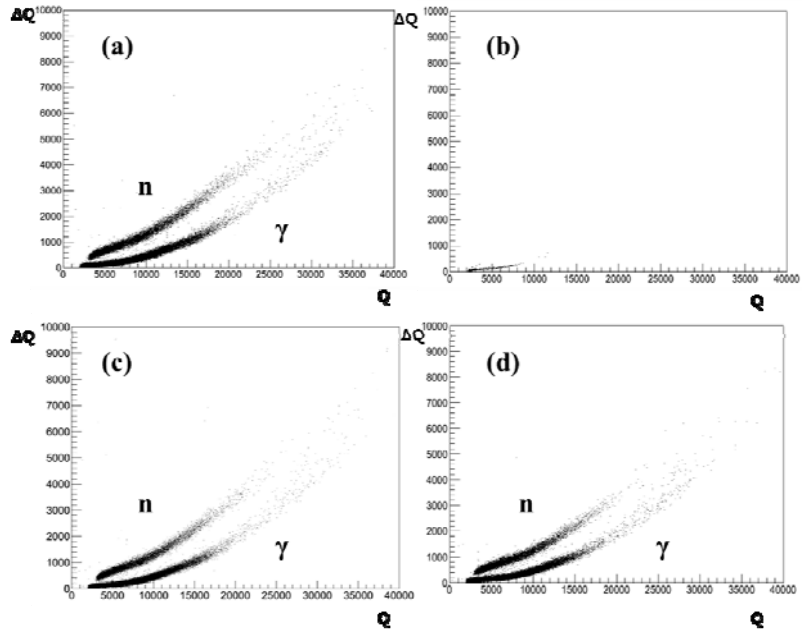

**Figure 10.** Neutron/gamma-ray scattered plot developed by using digital charge comparison method for $^{252}$Cf neutron source under the magnet fields: (a) no magnet, no magnetic shielding; (b) 1000 Gauss magnetic, no magnetic shielding; (c) 1000 Gauss magnetic, with 10 mm soft iron magnetic shielding; and (d) 4000 Gauss magnetic, with 10 mm soft iron magnetic shielding.

## 4. First measurement of neutron emission in KSTAR tokamak

The neutron diagnostic system was applied to experiments during the KSTAR campaign. The neutron diagnostic system instrument is comprised of a three-channel array located about 2 m away from the tokamak window, as shown in figure 11. The neutron detector system is composed of three scintillators: two differently sized stilbene scintillators ($\Phi$ 50 mm × 30 mm, $\Phi$ 50 mm × 40 mm) and a BC-501A liquid scintillator ($\Phi$ 50 mm × 50 mm).

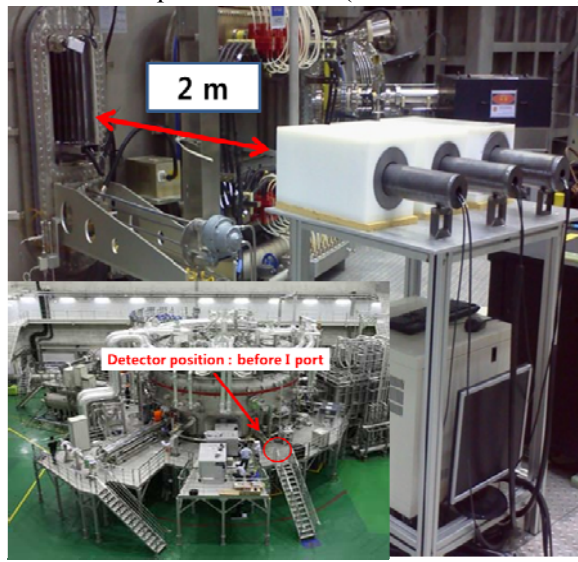

**Figure 11.** Location of neutron diagnostic system in KSTAR tokamak.

Figure 12 shows the waveforms of a typical Ohmic shot (plasma only). The plasma current is about 0.6 MA, and the electron line density is above $3.0 \times 10^{19}$ m$^{-2}$. The ion temperature measurement was not available after 5.0 s. The stilbene neutron diagnostic system observed neutrons from 1.5 to 4.6 s. The average neutron count rate was about 0.5 counts/ms for the Ohmic shots.

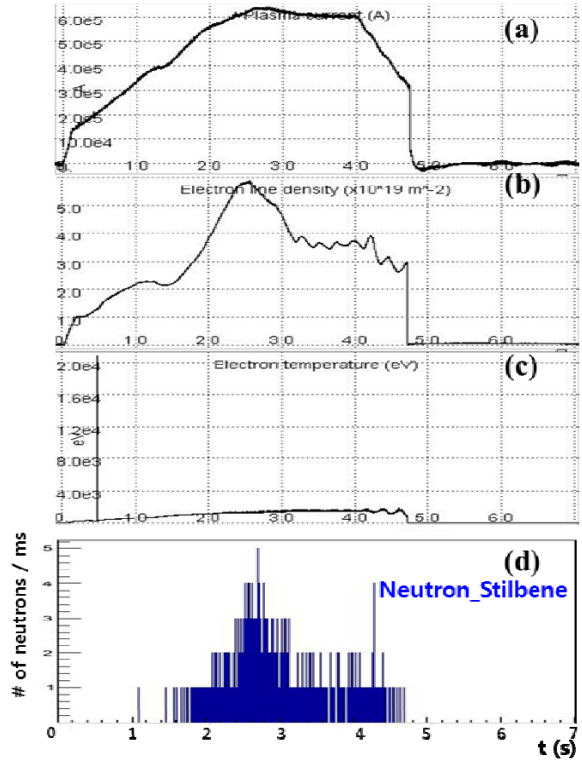

**Figure 12.** Waveforms of Ohmic shot: (a) plasma current (A), (b) electron line density ($10 \times 10^{19}$ m$^{-2}$), (c) electron temperature (eV), and (d) neutron signal for the stilbene diagnostic system.

Figure 13 shows the waveforms of an H-mode plasma. The plasma current was about 0.6 MA, and the electron line density was above $3.0 \times 10^{19}$ m$^{-2}$. No significant instabilities were observed in the core plasma region throughout this discharge. The ion temperature measurement was not available after 6.0 s. Neutron signals during the neutral beam injection are shown in figure 13(e). The average neutron count rate was about 50 counts/ms for the H-mode plasma shots. For the H-mode plasma shots, the neutron count time evolution of the stilbene neutron diagnostic system agreed well with the fission chamber time evolution, as shown in figures 12(d) and (e).

Figure 14 show a 2.45 MeV fusion neutron recoil proton spectrum and two-dimensional scattered plot of the partial charge versus total charge. The 2.45 MeV recoil protons were analyzed by using the light yields of mono-energetic neutrons and gamma radiation sources.

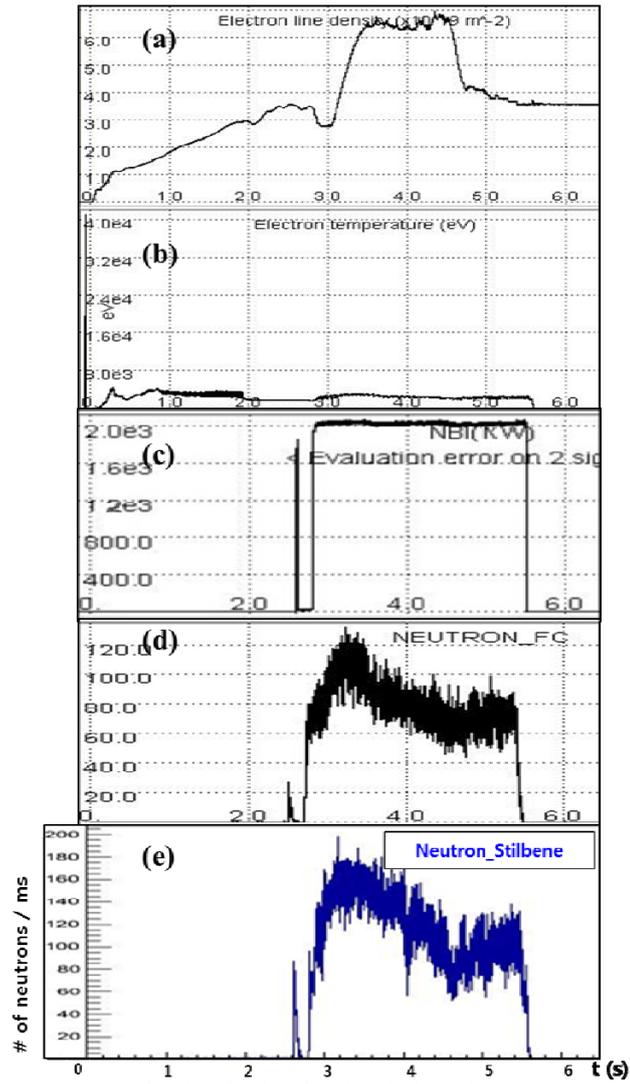

**Figure 13.** Waveforms of H-mode plasma: (a) electron line density ($10 \times 10^{19}$ m$^{-2}$), (b) electron temperature (eV), (c) total beam injection power, (d) neutron signal for fission chamber, and (e) neutron signal for stilbene diagnostic system.

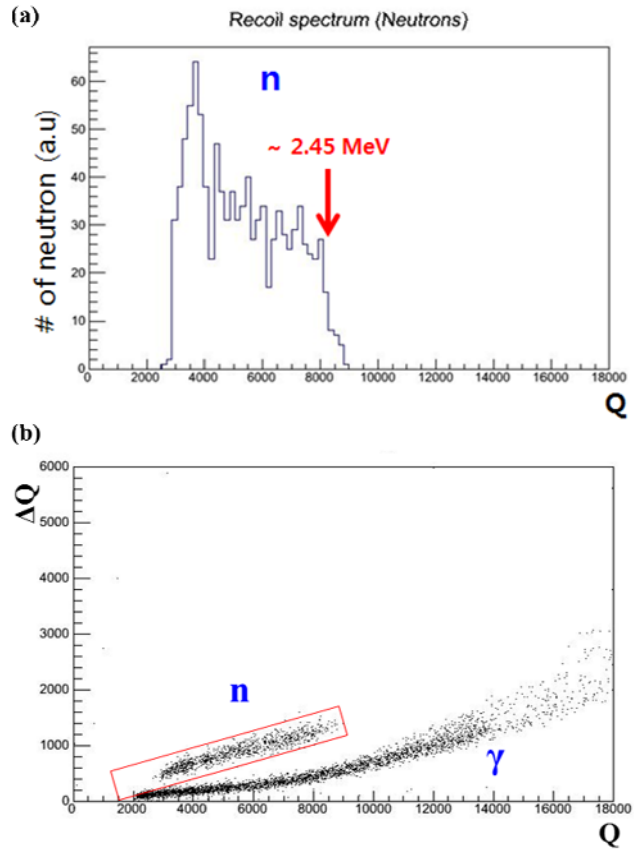

**Figure 14.** Recoil proton spectra with stilbene neutron diagnostic system in 2011 KSTAR campaign: (a) neutron recoil proton spectrum and (b) two-dimensional spectrum of partial charge versus total charge.

## 5. Conclusion

In conclusion, a stilbene Neutron diagnostic system was developed for neutron flux monitoring, spectrometry, and evaluation of the plasma performance of KSTAR. The neutrons and gamma-rays were successfully separated with the stilbene neutron diagnostic system. To evaluate the properties of the stilbene neutron diagnostic system, the magnetic shielding efficiency of 10 mm thick soft iron and detection efficiency of fast neutrons was tested experimentally using a $^{252}$Cf neutron source. The designed and fabricated stilbene neutron diagnostic system was found to show a good performance when measuring neutrons.

As the angle becomes smaller, the neutron path length of the collimator grows longer. Thus, the number of detected counts and intrinsic efficiency were confirmed to be lower. We calculated that the shielding efficiency of the diagnostic system used in this experiment was over 50%. The relative efficiency of the neutron diagnostic system using magnetic shielding of 10 mm soft iron was above 97% in a 1000 Gauss magnet field.

In the KSTAR tokamak, the first attempt to measure the recoiled proton spectrum was successful during the 2011 KSTAR campaign. The maximum count rate of this system was up to $2 \times 10^5$ neutrons/s, and a temporal resolution of 1 ms was obtained.

In future studies, system problems that were encountered will be addressed in the next KSTAR campaign. Under the NBI heating condition, data on the diagnostic system showed spread points that seemed to be due to a high count rate. Therefore, the count rates should be reduced by adjusting the distance between the tokamak window and our neutron diagnostic system. Also, additional high μ-metal shielding PMT will be considered to minimize the effects of magnetic fields. Supplementary magnetic shielding and signal line shielding are also required. Finally, a multi-detector array is being developed to measure the neutron profile in KSTAR.

## Acknowledgments

This work was supported by the nuclear R&D program of MEST, Korea (20110018425) and National R&D Program through the National Research Foundation of Korea (NRF) funded by the Ministry of Education, Science and Technology (20110018734).

## References

[1] Ishikawa M, Nishitani T, Morioka A, Takechi M, Shinohara K, Shimada M, Miura Y, Nagami M, and Kaschuck Yu A, *Rev. Sci. Instrum.* **73** (2002) 4237

[2] Denisov S, Dickey J, Dzierba A, Gohn W, Heinz R, Howell D, Mikels D, O'Neill D, Samoylenko V, Scott E, Smith P, and Teige S, *Nucl. Instrum. Meth.* **522** (2004) 467.

[3] Heltsley J.H et al., *Nucl. Instrum. Meth.* **A 263** (1988) 441.

[4] Knoll G.F, *Radiation detection and measurements*, John Willey and Sons, Inc., New York 2000.

[5] He Z, Bird A.J, Ramsden D, *Nucl. Instrum. Meth.* **336** (1993) 236

[6] Arulchakkaravarthi A, Santhanaraghavan P, Kumar R, Muralithar S, Ramasamy P, Nagarajan T, and Lan CW, *Mat. Chem. Phys.* **77** (2002) 77.

[7] Krainov IP, Galanov NZ, and Budakovsky SV, *Cryst. Res. Techn.* **24** (1989) 193.

[8] Leo WR, *Techniques for Nuclear and Particle Physics Experiments, A How-to Approach, Second Revised Edition,* Springer-Verlag, New York, Berlin, Heidelberg, 1994.

[9] Sysoeva E, Tarasox V, Zelenskaya O, *Nucl. Instrum. Meth.* **A 486** (2002), 67.